\documentclass[twocolumn]{aastex631}

\usepackage{amsmath}
\usepackage{xcolor}
\usepackage{graphicx}
\usepackage{hyperref}
\usepackage{booktabs}
\usepackage{multirow}
\usepackage{float}
\usepackage{placeins}
\usepackage[normalem]{ulem}

\newcommand{\sbout}[1]{\bgroup\markoverwith{\textcolor{violet}{\rule[.5ex]{2pt}{.4pt}}}\ULon{#1}}

\begin{document}

\title{Quantifying systematic uncertainties in white dwarf cooling age determinations}

\author[0009-0005-9976-1882]{Praneet Pathak}
\author[0000-0002-9632-1436]{Simon Blouin}
\author[0000-0001-8087-9278]{Falk Herwig}
\affiliation{Department of Physics \& Astronomy, University of Victoria, Victoria, BC, V8W 2Y2, Canada}

\begin{abstract}
Cooling ages of white dwarfs are routinely determined by mapping effective temperatures and masses to ages using evolutionary models. Typically, the reported uncertainties on  cooling ages only consider the error propagation of the uncertainties on the spectroscopically or photometrically determined $T_{\rm eff}$ and mass. However, cooling models are themselves uncertain, given their dependence on many poorly constrained inputs. This paper estimates these systematic model uncertainties. We use \texttt{MESA} to generate cooling sequences of $0.5-1.0\,M_{\odot}$ hydrogen-atmosphere white dwarfs with carbon--oxygen cores under different assumptions regarding the chemical stratification of their core, the thickness of their helium envelope, their hydrogen content, and the conductive opacities employed in the calculations. The parameter space explored is constrained by the range of values predicted by a variety of stellar evolution models and inferred from asteroseismological studies. For a $0.6\,M_{\odot}$ white dwarf, we find an uncertainty of 0.03\,Gyr at $10{,}000\,$K (corresponding to a 5\% relative uncertainty) and 0.8\,Gyr at $4000\,$K (9\%). This uncertainty is significant, as it is comparable to the age uncertainty obtained by propagating the measurement errors on $T_{\rm eff}$ and mass for a typical white dwarf. We also separately consider the potential impact of $^{22}$Ne shell distillation, which plausibly leads to an additional uncertainty of $\sim 1\,$Gyr for crystallized white dwarfs. We provide a table of our simulation results that can be used to evaluate the systematic model uncertainty based on a white dwarf's $T_{\rm eff}$ and mass. We encourage its use in all future studies where white dwarf cooling ages are measured.
\end{abstract}


\section{Introduction}
\label{sec:intro}
White dwarfs represent the final evolutionary stage for the vast majority of stars. Having exhausted their nuclear fuel, they enter a long cooling phase that is relatively simple compared to earlier stages of stellar evolution \citep{Mestel1952,Fontaine2001,Althaus2010}. This cooling process spans billions of years and, if accurately modeled, can allow white dwarfs to be used as cosmic clocks. White dwarfs have been used to estimate the ages and star formation histories of various stellar populations \citep{Winget1987,Garcia-Berro2010,Kalirai2012,Kilic2017,Fantin2019,Isern2019,Cukanovaite2023}, constrain the Hubble constant \citep{Cimatti2023}, and date ancient planetary systems \citep{Hollands2018,Kaiser2021,Blouin2022,Elms2022}. However, our understanding of both the late stages of stellar evolution leading to white dwarf formation and the subsequent white dwarf cooling process remains incomplete, which limits our ability to accurately determine white dwarf cooling ages.

A first major source of uncertainty is the core composition profile of white dwarfs. This ratio is determined by the helium-burning phases of the progenitor star, where three key uncertainties come into play. First, the $^{12}{\rm C}(\alpha,\gamma)^{16}{\rm O}$ reaction rate is poorly constrained, leading to variations of about $\pm 0.1$ in the central oxygen mass fraction \citep{DeGeronimo2017,Chidester2022,Pepper2022}. Second, the physics of convective boundary mixing during helium burning is not well understood, affecting both the final carbon/oxygen ratio and the size of the homogeneous core region \citep{Straniero2003,Constantino2015,Salaris2017,Giammichele2022,Blouin2024}. Third, the thermal pulse phase on the asymptotic giant branch introduces additional complexity, as the uncertain number of pulses influence the chemical profile of the outer region of the core \citep{Herwig2000,Weiss2009,DeGeronimo2017}. These uncertainties in the core composition profile have important implications for white dwarf cooling ages. The composition affects the total heat content of the core for a given mass, and it influences the crystallization process, both in terms of when it begins and how much energy it releases \citep{Althaus2012,Blouin2020b,Bauer2023}.

The envelope composition is a second significant source of uncertainty in modeling white dwarf evolution. The hydrogen envelope mass is especially uncertain. Both asteroseismology and spectral evolution studies point to a wide intrinsic scatter in the white dwarf population, ranging from $M_{\rm H}/M_{\star} \sim 10^{-4}$ to $10^{-16}$ \citep{Castanheira2009,Rolland2018,Giammichele2022,Bedard2024}. This spread is understood to be the result of diverse evolutionary paths: some stars undergo very late thermal pulses where nearly all hydrogen is burnt; others experience late thermal pulses resulting in partial hydrogen depletion; and many evolve without such events, retaining their original thick hydrogen envelopes \citep{Althaus2005b,Althaus2005}. The helium layer mass is better constrained, with a typical value of $M_{\rm He}/M_{\star} \sim 10^{-2}$ \citep{Renedo2010,Giammichele2022}, although this is also dependent on the number of thermal pulses. Importantly, the envelope significantly impacts white dwarf cooling rates by regulating the flow of energy from the degenerate core to the surface, with thicker hydrogen layers generally leading to slower cooling.

Beyond uncertainties in composition, there are also significant open questions regarding the physics of white dwarf interiors that impact cooling rates. One major area of uncertainty lies in the conductive opacities, which play a crucial role in energy transport within white dwarfs. While the conductive opacities in the strongly degenerate, strongly coupled carbon--oxygen cores are well understood, the regime of partial degeneracy and moderate coupling found in the helium and hydrogen layers presents challenges \citep{Saumon2022}. Recent work has led to revised calculations of conductive opacities in this partially degenerate regime \citep{Blouin2020}. These new opacities differ from previous values \citep{Cassisi2007} in a narrow region of intermediate degeneracy, potentially leading to faster cooling, especially for massive white dwarfs. However, there is ongoing debate about the accuracy of these new calculations \citep{Cassisi2021}. This is an unresolved issue that translates into a significant source of uncertainty for white dwarf cooling \citep{Salaris2022}.

Another important question for white dwarf cooling is the possibility of $^{22}$Ne distillation. This process, first proposed by \cite{Isern1991}, involves the separation of $^{22}$Ne impurities (or other neutron-rich impurities, \citealt{Shen2023}) from the background carbon--oxygen plasma during crystallization, leading to the release of gravitational energy that can significantly delay cooling. Recent work has shown that this mechanism likely explains the puzzling ``Q branch'' feature in the Gaia color--magnitude diagram, an overdensity of high-mass white dwarfs experiencing multi-gigayear cooling delays \citep{GaiaCollaboration2018,Tremblay2019,Cheng2019,Blouin2021L,Shen2023,Bedard2024N}, as well as the faint end of the observed luminosity function of the old metal-rich open cluster NGC~6791 \citep{Salaris2024}. While we have convincing evidence that distillation occurs in some white dwarfs (i.e., many Q-branch objects and the $^{22}$Ne-rich white dwarfs of NGC~6791), its prevalence and impact in the general white dwarf population remains uncertain. Theoretical calculations suggest that it could cause cooling delays of 0.5--2.0\,Gyr in typical white dwarfs descending from solar-metallicity progenitors \citep{Segretain1996,Blouin2020b,Blouin2021L}. For these stars, distillation would occur in a thin shell around an already solid central core rather than at the center of the star, as is thought to be the case for Q-branch objects and the $^{22}$Ne-rich white dwarfs in NGC~6791. However, observational confirmation of this shell distillation mechanism is still lacking \citep{Venner2023,Barrientos2024}. The magnitude of any such delay would depend sensitively on the initial core composition.

The current standard practice for determining white dwarf ages involves measuring their effective temperature and mass through spectroscopic or photometric methods, and then mapping these parameters to an age using pre-calculated cooling tracks \citep[such as those provided by][]{Renedo2010,Camisassa2019,Bedard2020,Salaris2022}. The resulting cooling age is typically reported with an error bar that reflects only the propagation of uncertainties in the inferred $T_{\rm eff}$ and mass \citep{Kiman2022}. However, the systematic errors inherent in the cooling tracks themselves are almost always overlooked. This paper aims to address this gap by quantifying the systematic uncertainties in white dwarf cooling models and presenting the results in a format that can be easily incorporated into future works.

It is important to clarify that our focus in this paper is specifically on cooling ages, rather than total stellar ages. For most white dwarfs, the dominant source of error in total age estimates stems from uncertainties in the initial-final mass relation (IFMR, \citealt{Cummings2018,Cunningham2024,Hollands2024}). While we do not address IFMR uncertainties in this work, we note that this source of error can be largely mitigated by focusing on more massive white dwarfs, which have much shorter main-sequence lifetimes \citep{Isern2019}.

The structure of this paper is as follows. Section~\ref{sec:methods} outlines our methodology for quantifying systematic uncertainties in white dwarf cooling ages using \texttt{MESA} simulations. Section~\ref{sec:results} presents our results and provides our recommended uncertainties for future use. Finally, we conclude in Section~\ref{sec:conclusion}.

\section{Methods}
\label{sec:methods}
We employ the Modules for Experiments in Stellar Astrophysics \citep[\texttt{MESA},][]{MESA1,MESA2,MESA3,MESA4,MESA5} to simulate white dwarf cooling. We use \texttt{MESA} version \texttt{r23.05.1}, which incorporates carbon--oxygen phase separation in crystallizing white dwarfs \citep{Bauer2023}. We generate cooling sequences for white dwarfs spanning a mass range of 0.5 to 1.0$\,M_{\odot}$, in increments of 0.1\,$M_{\odot}$. For each mass, we systematically vary both the composition profile and conductive opacities to factor in the uncertainties outlined in Section \ref{sec:intro}.

\subsection{Composition profile parametrization}
To account for the compositional uncertainties in white dwarf interiors, we use the MESA tool \texttt{wd\_builder}\footnote{\url{https://github.com/MESAHub/mesa-contrib}} to construct initial white dwarf models with varied chemical composition profiles. The core oxygen composition profile is parameterized using an eight-parameter Akima spline profile, following the approach of \citet{Giammichele2017}. This Akima spline parametrization was chosen for its flexibility and ability to produce smooth profiles. It can reproduce a wide range of core oxygen profiles, from simple single-transition profiles to more complex profiles with two distinct transitions, allowing it to approximate core structures predicted by various evolutionary models. The use of Akima splines, as opposed to other interpolation methods like cubic splines, avoids artificial oscillations in the resulting profiles.

\begin{figure}
   \centering
   \includegraphics[width=\columnwidth]{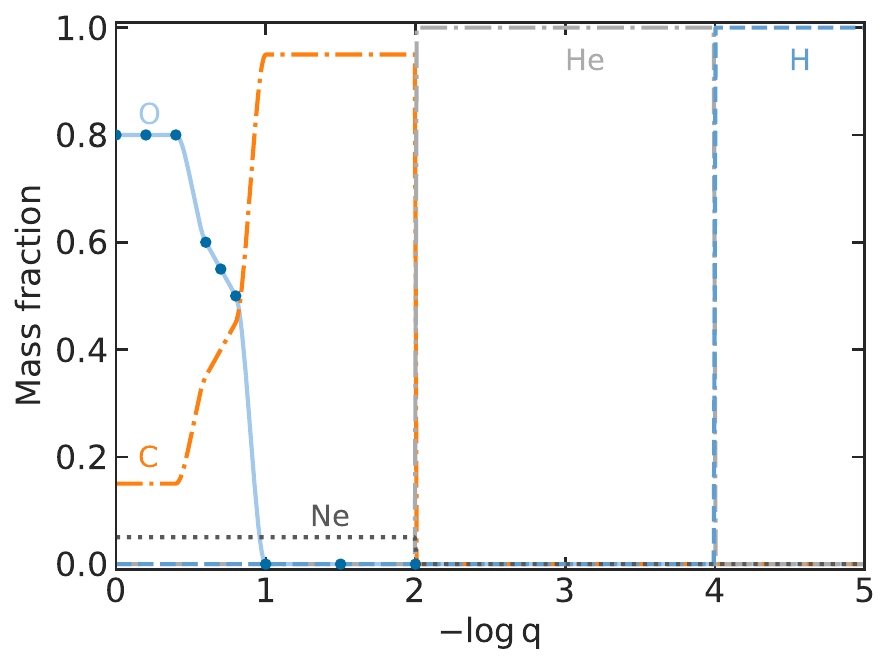}
      \caption{Representative chemical composition profile of a white dwarf model used as an input for \texttt{wd\_builder}. The profile is defined by eleven parameters, including eight for the oxygen profile which is constructed using Akima splines \citep{Giammichele2017}. The nine dots show the oxygen abundance at specific mass coordinates used to define the splines. The x-axis shows the logarithmic mass coordinate, where $\log q = \log \left(1 - m(r)/M_\star\right) = 0$ represents the center of the star.}
         \label{fig:composition}
   \end{figure}

The complete chemical composition profile includes eleven parameters. In addition to the oxygen profile, we include a constant amount of $^{22}$Ne in the core, with the remainder being $^{12}$C. Outside the core, we add a helium envelope of thickness $M_{\rm He}/M_{\star}$ and a thin hydrogen layer of thickness $M_{\rm H}/M_{\star}$. A typical profile is shown in Figure~\ref{fig:composition}.

\subsection{Parameter space constraints}
\begin{figure}
    \centering
    \includegraphics[width=\columnwidth]{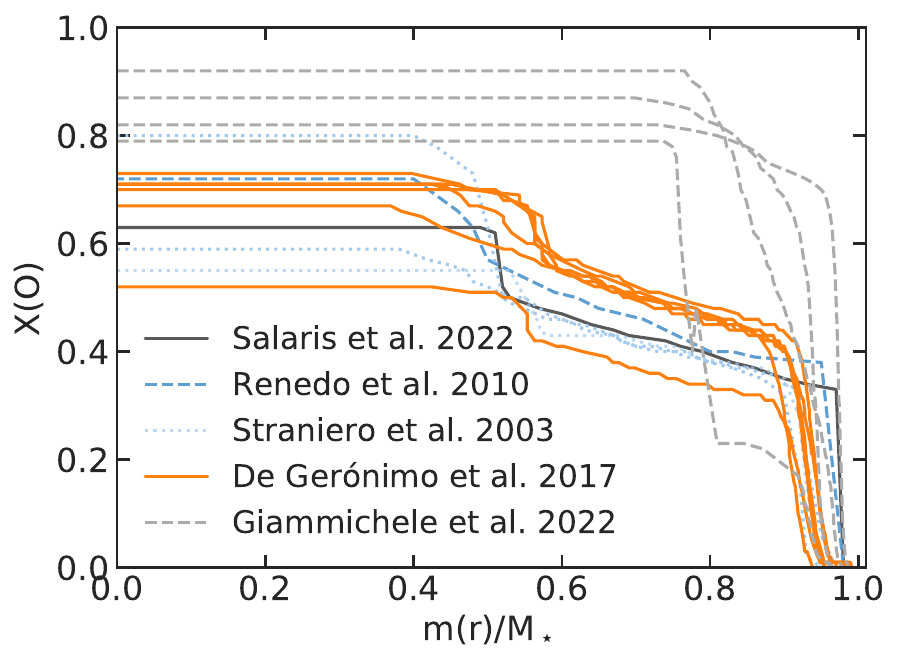}
    \caption{Oxygen abundance profiles for $\simeq 0.6 \, M_\odot$ white dwarf from various evolutionary models and asteroseismological studies. These profiles were used to inform the parameter space sampled in this work.
}
    \label{fig:input_Oprofiles}
\end{figure}

To inform the parameter space for our simulations, we examined model predictions from various stellar evolution codes and inferences from asteroseismology and fitted the Akima spline model to their oxygen profiles. More specifically, we considered the oxygen profiles of the four DA white dwarfs analyzed by \cite{Giammichele2022}, the models shown in \cite{Straniero2003}\footnote{We eliminated the inversions visible in the oxygen abundance profiles of \cite{Straniero2003} using the chemical rehomogenization method of \cite{Salaris1997}.}, the \texttt{BASTI} models of \cite{Salaris2022}, and the \texttt{LPCODE} models of \cite{Renedo2010} and \cite{DeGeronimo2017}. These profiles are shown in Figure~\ref{fig:input_Oprofiles}.

From these fits, we defined the parameter space for our core oxygen profile. For each of the eight Akima spline parameters, we determined the minimum and maximum values obtained from fitting this diverse set of models and asteroseismological inferences. We then defined a uniform distribution spanning from this minimum to maximum for each parameter. This approach allows us to capture the range of predictions from current evolutionary calculations and asteroseismological analyses, reasonably encompassing the uncertainties in core oxygen profiles. The ranges of values explored for each parameter are given in Table~\ref{tab:param_range}.\footnote{While we have not explicitly included \texttt{MESA} models in the building of these distributions, we verified that they encompass the \texttt{MESA} models of \cite{Bauer2023}.} For simplicity, we use the same range of values for the core oxygen parameters across all white dwarf masses. This approach is required due to the fragmentary nature of available constraints, as published models and asteroseismology data exist only for specific masses and a handful of stars. The meaning of each parameter is defined in \citealt{Giammichele2017} (see their Figure~1); as examples, $t_1 - \Delta t_1$ represents the size of the inner homogeneous core (in terms of $\log q$) and $X_{\rm center} ({\rm O})$ is the oxygen abundance in that same region. 

\vspace{-10pt}
\begin{deluxetable*}{r|ccccccc}
\tablecaption{Range of values used to parameterize the white dwarf model composition profiles\label{tab:param_range}}
\tablewidth{0pt}
\tablehead{
\colhead{Parameter\tablenotemark{a}} & \multicolumn{6}{c}{White dwarf mass ($M_\odot$)} \\
\cline{2-7}
\colhead{} & \colhead{0.5} & \colhead{0.6} & \colhead{0.7} & \colhead{0.8} & \colhead{0.9} & \colhead{1.0}
}
\startdata
$X_{\rm center} ({\rm O})$ & \multicolumn{6}{c}{$U(0.50, 0.90)$} \\
$t_1$ & \multicolumn{6}{c}{$U(0.30, 0.80)$} \\
$\Delta t_1$ & \multicolumn{6}{c}{$U(0.00, 0.25)$} \\
$t_1 (\rm{O})$ & \multicolumn{6}{c}{$U(0.25, 0.75)$} \\
$t_2$ & \multicolumn{6}{c}{$U(1.05, 1.60)$} \\
$\Delta t_2$ & \multicolumn{6}{c}{$U(0.00, 0.25)$} \\
$t_2 (\rm{O})$ & \multicolumn{6}{c}{$U(0.20, 0.70)$} \\
$X_{\rm envl} ({\rm O})$ & \multicolumn{6}{c}{$U(0.00, 0.40)$} \\
$X_{\rm core} (^{22}{\rm Ne})$ & \multicolumn{6}{c}{$0.014 \times {\rm LogNormal}(\mu = -0.2, \sigma = 0.5)$} \\
$-\log q_{\rm He}$ & $U(1.50, 3.00)$ & $U(1.50, 3.00)$ & $U(1.50, 3.00)$ & $U(1.80, 3.00)$ & $U(2.10, 3.00)$ & $U(2.35, 3.00)$ \\
$-\log q_{\rm H}$ & $U(4.00, 6.00)$ & $U(4.00, 6.00)$ & $U(4.10, 6.00)$ & $U(4.30, 6.00)$ & $U(4.75, 6.00)$ & $U(4.75, 6.00)$ \\
\enddata
\tablenotetext{a}{We refer the reader to \citet{Giammichele2017} for a definition of the first eight parameters, which encode the core composition profile. The last two parameters define the thickness of the helium and hydrogen layers, where $q = 1 - m(r)/M_\star$.}
\tablecomments{$U(a,b)$ denotes a uniform distribution between $a$ and $b$.}
\end{deluxetable*}
\vspace{-15pt}

\begin{figure}
    \centering
    \includegraphics[width=\columnwidth]{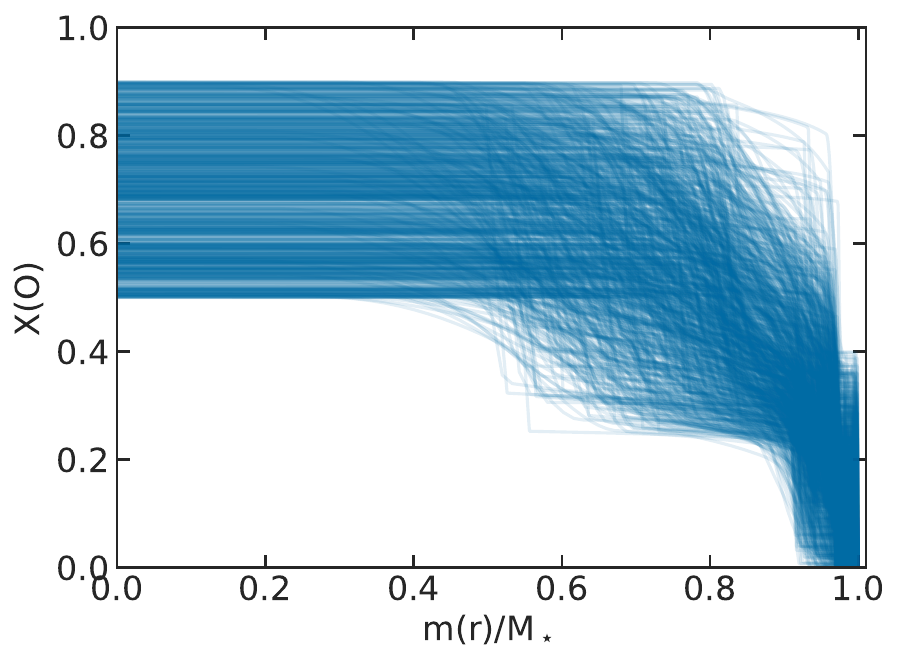}
    \caption{Oxygen abundance profiles generated by randomly sampling the distributions that we defined for each of the eight Akima spline parameters (Table~\ref{tab:param_range}).
}
    \label{fig:comp_examples}
\end{figure}

Figure~\ref{fig:comp_examples} shows examples of oxygen abundance profiles generated from these eight distributions of Akima spline parameters. Each profile corresponds to a distinct model that we then evolve with \texttt{MESA}. Note that to be consistent with standard stellar evolution model predictions we have constrained our sampling of the parameter distributions to enforce that the oxygen abundance is monotonically decreasing from the center outward. For example, this means that once the central oxygen abundance $X_{\rm center}({\rm O})$ is defined, all other parameters are constrained such that the oxygen abundance cannot increase above this central value at any point in the profile. 

We assume a constant abundance $^{22}$Ne trace abundance in the core. For each model, the progenitor's metallicity $\left[ {\rm Fe/H} \right]$ is selected from a normal distribution with $\mu=-0.2$ and $\sigma=0.5$ (an appropriate prior for disk white dwarfs, \citealt{Moss2022}), and $X_{\rm core} (^{22}{\rm Ne})$ is assumed to correspond to $0.014 \times 10^{\left[ {\rm Fe/H} \right]}$.
   
For the envelope composition, we vary the helium and hydrogen layer masses within ranges again informed by evolutionary calculations and observational constraints \citep{Renedo2010,Romero2013,Giammichele2022}. For both these parameters, the true probability distributions remain unknown, and we therefore adopt uniform distributions for simplicity. The upper limits for these envelope masses are set to ensure the model does not undergo nuclear burning \citep[as expected for the vast majority of white dwarfs,][]{Althaus2015}, with specific limits depending on the white dwarf mass (Table~\ref{tab:param_range}).

In this work, we limit our analysis to white dwarfs with hydrogen layers thick enough to maintain a pure hydrogen atmosphere throughout their evolution ($M_{\rm H}/M_{\star} \gtrsim 10^{-6}$, \citealt{Rolland2018}). This represents approximately 75\% of the white dwarf population \citep{Bedard2024}. These white dwarfs would generally be classified as DA (showing hydrogen absorption lines) or DC (featureless spectra) at lower temperatures. However, it is important to note that not all observed DA or DC white dwarfs have such thick hydrogen layers.

\subsection{Constitutive physics}
Given current uncertainties on the envelope conductive opacities, we use the \cite{Cassisi2007} conductivities for half of our models and the \cite{Blouin2020} for the other half. The atmosphere boundary conditions are based on the \cite{Rohrmann2011} tables, appropriate for the hydrogen-atmosphere white dwarfs considered here. $^{22}$Ne gravitational settling was omitted from our calculations to reduce computational time. Our tests with a limited number of models showed that including this process had a negligible effect on the overall cooling age uncertainties. Templates of our \texttt{MESA} inlists can be retrieved at the DOI \href{https://doi.org/10.5281/zenodo.13831121}{10.5281/zenodo.13831121}.

The final uncertainty we consider relates to $^{22}$Ne shell distillation, a process which may significantly delay white dwarf cooling but is not included in standard evolutionary models. To account for this uncertainty while maintaining compatibility with existing cooling models, we adopt a two-component approach. First, we calculate a baseline systematic uncertainty using our ensemble of models without distillation, capturing the spread due to variations in core composition, envelope masses, and conductive opacities. To quantify the potential impact of shell distillation, we create a second ensemble by applying a cooling delay to all of our original models. The magnitude of this delay is drawn from a uniform distribution between 0.5 and 2~Gyr for each model, consistent with theoretical predictions \citep{Segretain1996,Blouin2020b,Blouin2021L}. The effective temperature range in which this delay is introduced is determined using both theoretical considerations and observational evidence. Shell distillation is expected to occur after the white dwarf has begun crystallizing but before crystallization is complete. \cite{Blouin2021L} suggested that shell distillation takes place when $\sim 60$\% of the core is crystallized assuming a uniform $X({\rm O})=0.60$ and $X(^{22}{\rm Ne}) = 0.014$ core composition. To account for uncertainties in core composition \citep[which determines the $T_{\rm eff}$ at which shell distillation is expected to start,][]{Blouin2021L}, we assume that the additional delay is injected between the effective temperatures corresponding to 35\% and 65\% crystallization on average in our cooling models. This is consistent with the observational mass--$T_{\rm eff}$ diagram presented by \cite{Kilic2024}, which shows an overdensity of white dwarfs in certain temperature ranges at high mass. This feature cannot be fully explained by standard crystallization cooling delays and may provide evidence for $^{22}$Ne distillation. 

% \begin{figure}
%     \centering
%     \includegraphics[width=\hsize]{figures/distillation_overdensity-new.pdf}
%     \caption{White dwarf mass--$T_{\rm eff}$ distribution from \cite{Kilic2024}. The orange and black lines mark the locations where white dwarfs of a given mass are 35\% and 65\% crystallized, respectively, based on our average cooling models. These lines bracket the temperature range where we inject the $^{22}$Ne shell distillation delay in our models. Note that the completeness of the \cite{Kilic2024} sample drops sharply below 5000\,K.}
%     \label{fig:distillation_overdensity}
% \end{figure}

We calculate the average cooling times for both ensembles of models (with and without shell distillation) and determine the mean delay ($Y$) by comparing these averages for different masses and effective temperatures. This two-component approach allows us to report cooling age uncertainties in the format $\pm X (+Y)$~Gyr, where $X$ is the systematic uncertainty from the composition and conductive opacities, and $Y$ is the potential delay due to distillation. This method clearly delineates the effect of distillation, facilitating future adjustments as our understanding improves, while providing a practical way to incorporate these uncertainties into existing models without distillation delays.

\subsection{Summary}
\begin{figure}
    \centering
    \includegraphics[width=\hsize]{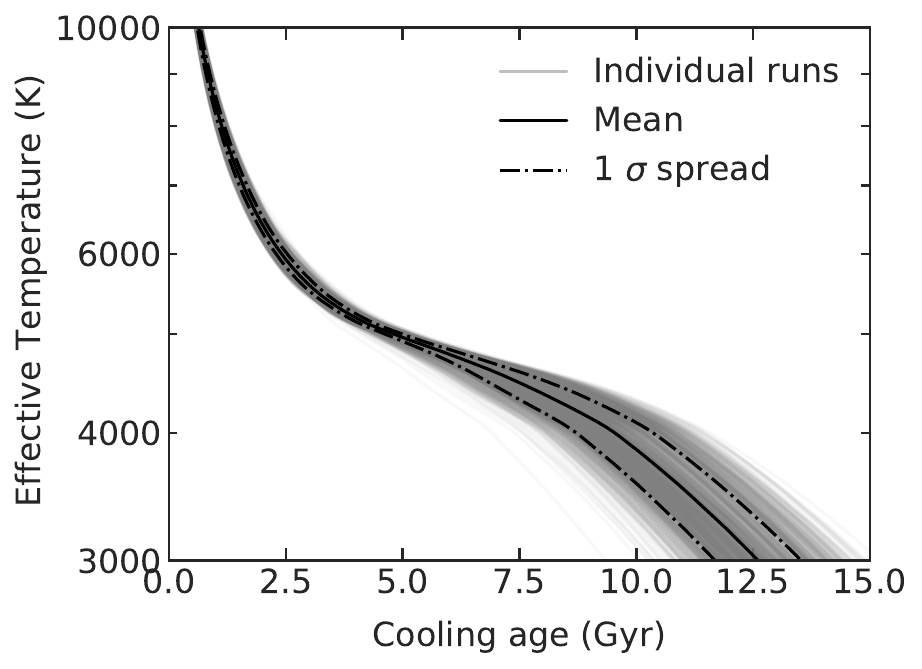}
    \caption{Distribution of cooling tracks for a 0.6\,$M_{\odot}$ white dwarf. The gray lines represent individual cooling sequences from our 1000 model runs, each with different combinations of core and envelope compositions and conductive opacities. The solid black line shows the mean cooling track, while the dotted lines indicate the 1$\sigma$ spread around this mean. $^{22}$Ne distillation is omitted for these cooling tracks.}
    \label{fig:cooling_tracks}
\end{figure}

To summarize our approach, we vary a total of 12 parameters for each white dwarf model: (1) eight parameters defining the oxygen abundance profile in the core, (2) the $^{22}$Ne abundance in the core, (3) the helium layer mass, (4) the hydrogen layer mass, and (5) the choice of conductive opacities. For each white dwarf mass considered (0.5, 0.6, 0.7, 0.8, 0.9, and 1.0$\,M_{\odot}$), we generate 1000 models by randomly sampling these 12 parameters. We then create two ensembles: one without $^{22}$Ne distillation and another with distillation-induced cooling delays applied. For the latter, we randomly determine the timing and duration of the delay for each model. This approach results in two distributions of cooling tracks for each mass, capturing the range of possible evolutionary paths. Figure~\ref{fig:cooling_tracks} illustrates the distribution of cooling tracks for a 0.6$\,M_{\odot}$ white dwarf without distillation, showcasing the spread in cooling behavior arising from our parameter variations. We use the 1$\sigma$ spread in cooling age at a given effective temperature in the no-distillation ensemble as our measure of the baseline systematic uncertainty ($X$). The mean difference in cooling age between the two ensembles at each effective temperature provides our estimate of the potential delay due to $^{22}$Ne distillation ($Y$), and we will report total uncertainties as $\pm X (+Y)$~Gyr.

\section{Results}
\label{sec:results}
\subsection{Uncertainties without distillation}

\begin{figure*}
\centering
\includegraphics[width=0.49\textwidth]{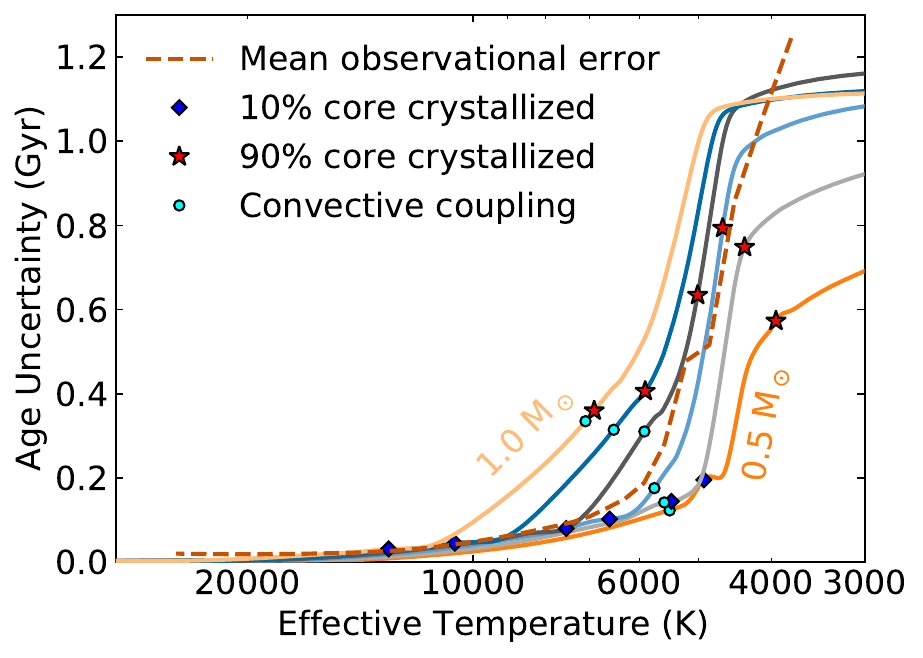}
\includegraphics[width=0.49\textwidth]{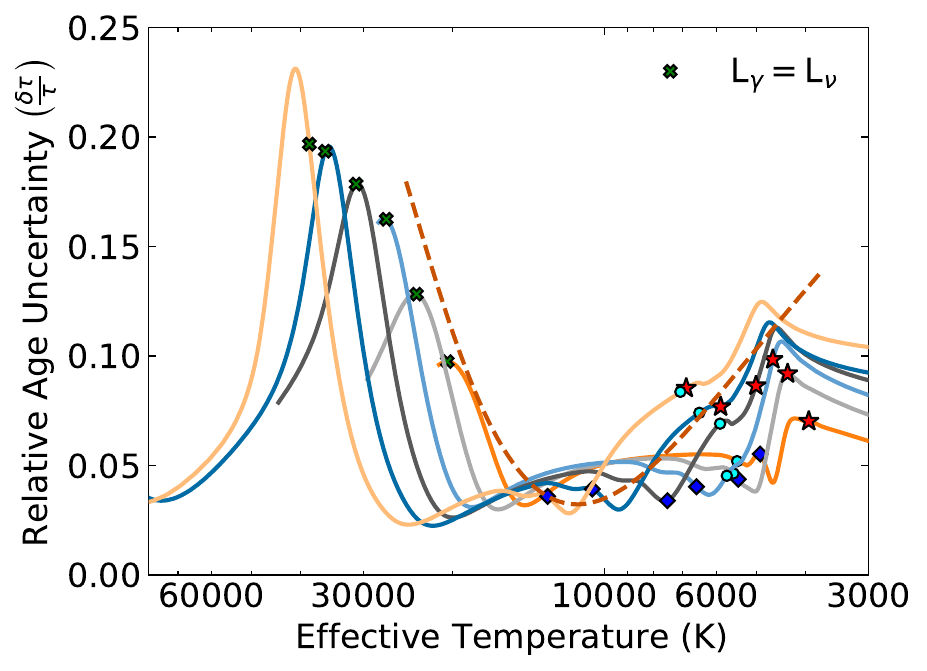}
\caption{\textit{Left:} Systematic uncertainty in white dwarf cooling ages as a function of effective temperature for masses ranging from 0.5 to 1.0$\,M_{\odot}$. Each curve represents a different mass, decreasing from left to right in 0.1$\,M_{\odot}$ increments. Key physical transitions are marked: blue diamonds indicate 10\% core crystallization, red stars show 90\% core crystallization, and cyan circles denote the onset of convective coupling between the core and envelope. We defined convecting coupling as occurring when the base of the superficial hydrogen convection zone first reaches the upper boundary of the degenerate core, located where the electron degeneracy parameter $\eta = 0$ \citep{Fontaine2001}. The orange dotted line represents the average age uncertainty based on error propagation of $T_{\rm eff}$ and mass measurements \citep{O'Brien2024}. \textit{Right:} Relative systematic uncertainty in white dwarf cooling ages as a function of effective temperature. Green crosses indicate where photon luminosity equals neutrino luminosity for each mass. All other symbols have the same meaning as in the left panel.}
\label{fig:age_uncert}
\end{figure*}

Figure \ref{fig:age_uncert} presents the systematic uncertainties in white dwarf cooling ages derived from our ensemble of models, when distillation delays are omitted. The left panel shows the absolute age uncertainty as a function of effective temperature for white dwarf masses ranging from 0.5 to 1.0\,$M_\odot$, while the right panel displays the relative uncertainty. Notably, the magnitude of these systematic uncertainties is comparable to, and in many cases exceeds, the typical observational uncertainties in cooling ages (orange dotted line), which we derived here from the error propagation of $T_\mathrm{eff}$ and mass measurements in the 40\,pc sample of \cite{O'Brien2024}. 

\begin{figure}
\centering
\includegraphics[width=\hsize]{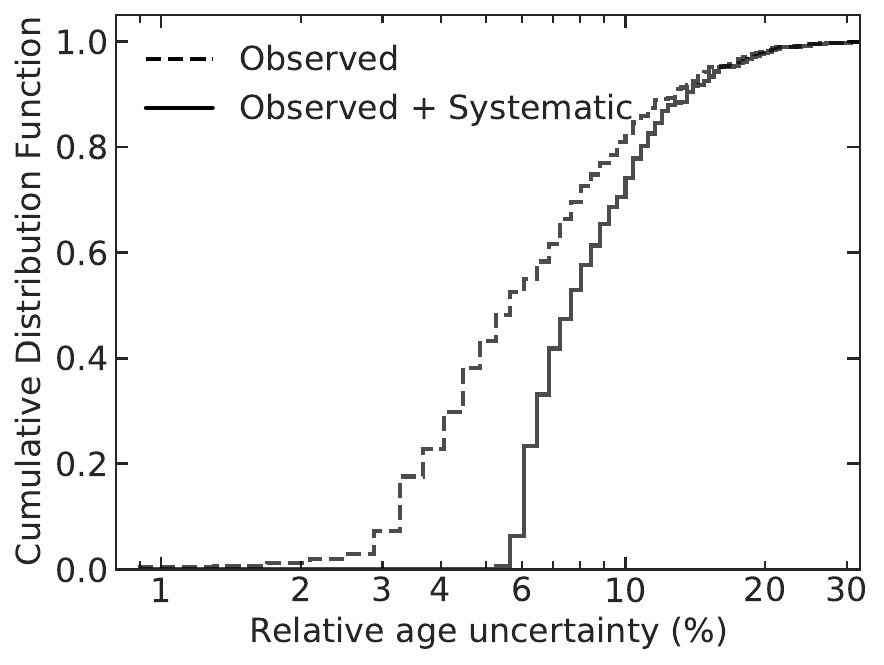}
\caption{Cumulative distribution function (CDF) of relative age uncertainties for white dwarfs in the \cite{O'Brien2024} sample. The dashed line represents the CDF considering only observational uncertainties in cooling ages. The solid line shows the CDF when both observational and systematic cooling model uncertainties are included. This comparison excludes potential additional uncertainties from $^{22}$Ne distillation-induced cooling delays, which would further increase the total uncertainty for a significant fraction of this sample.}
\label{fig:rel_age_uncert_histogram}
\end{figure}

The impact of these systematic uncertainties on the overall age uncertainty budget is further illustrated in Figure \ref{fig:rel_age_uncert_histogram}. This figure shows the cumulative distribution function (CDF) of relative age uncertainties for white dwarfs in the \cite{O'Brien2024} sample, both with and without the inclusion of our derived systematic uncertainties. The shift in the CDF when systematic uncertainties are incorporated (solid line) compared to observational uncertainties alone (dashed line) highlights the significant contribution of model uncertainties to the total error budget. Note that these uncertainties were added in quadrature, as systematic cooling model uncertainties are independent of observational errors on the measured $T_{\rm eff}$ and masses. 

The relative age uncertainty curves in Figure \ref{fig:age_uncert} (right panel) exhibit two notable features: a pronounced bump at high effective temperatures (between $40{,}000\,$K and $20{,}000\,$K depending on the mass), and a second one at lower temperatures just after 5000\,K. The high-temperature bump in relative age uncertainty coincides closely with the point where photon luminosity equals neutrino luminosity (green crosses in Figure \ref{fig:age_uncert}), indicating that this feature is primarily associated with neutrino cooling. During this phase, variations in initial cooling rates, caused for instance by differences in envelope thickness or conductivities \citep{Blouin2020}, lead to divergent core temperatures between the different models. Cooler cores then experience less efficient neutrino cooling, resulting in longer cooling timescales. While in relative terms the systematic uncertainty is very significant at these high temperatures, it remains negligible in absolute terms given the young ages of white dwarfs undergoing neutrino cooling.

Similarly, the increase in relative age uncertainty at lower temperatures is largely attributable to uncertainties in conductive opacities, particularly their impact after convective coupling. Convective coupling occurs when the superficial convection zone extends deep enough to reach the degenerate, conductive layers of the envelope (indicated by cyan circles in Figure \ref{fig:age_uncert}). This event, occurring around 6000\,K (with the exact temperature depending on the white dwarf's mass and envelope composition), marks the onset of strong thermal coupling between the core and outer layers \citep{Fontaine2001}. The process of convective coupling temporarily slows cooling as the star adjusts to this new energy transport regime. The magnitude of this cooling delay depends on the amount of excess thermal energy in the core at the onset of coupling. Models with higher envelope conductivities cool more efficiently prior to convective coupling, resulting in cooler cores when coupling occurs. Consequently, these models experience a smaller cooling delay during the convective coupling phase, leading to age divergences that become particularly pronounced below 6000\,K \citep{Blouin2020}.

\begin{deluxetable*}{ccccccc}
\tablecaption{Relative systematic cooling age uncertainty (\%) as a function of mass and effective temperature for hydrogen-atmosphere white dwarfs with carbon--oxygen cores
\label{tab:rel_age_uncert}}
\tablewidth{0pt}
\tablehead{
\colhead{$T_\mathrm{eff}$ (K)} & 
\colhead{0.5\,$M_\odot$} & 
\colhead{0.6\,$M_\odot$} & 
\colhead{0.7\,$M_\odot$} & 
\colhead{0.8\,$M_\odot$} & 
\colhead{0.9\,$M_\odot$} & 
\colhead{1.0\,$M_\odot$}
}
\startdata
20000 & 9.7  & 8.0 & 3.9 & 2.6 & 2.5 & 3.1 \\
17500 & 7.1  & 3.6 & 3.2 & 3.1 & 3.2 & 3.7 \\
15000 & 3.4  & 3.3 & 4.1 & 3.9 & 3.8 & 3.7 \\
12500 & 4.0  & 4.4 & 4.8 & 4.5 & 4.1 & 3.2 \\
10000 & 5.1  & 5.1 & 5.1 & 4.6 & 3.4 & 5.1 (+38.2) \\
9000 & 5.3  & 5.2 & 5.2 & 4.3 & 3.3 (+9.4)& 6.5 (+52.0) \\
8000 & 5.4  & 5.3 & 4.8 & 3.7 & 5.2 (+33.4)& 7.5 (+40.4) \\
7000 & 5.5  & 5.2 & 4.6 & 4.3 & 6.8 (+34.6)& 8.4 (+31.0)\\
6000 & 5.5  & 4.9 & 3.9 & 6.8 (+20.3)& 7.7 (+24.5)& 9.0 (+22.8)\\
5500 & 5.2  & 4.5 & 5.0 (+10.6)& 7.0 (+22.5)& 8.3 (+19.5)& 10.2 (+18.7)\\
5000 & 5.6  & 3.8 (+9.1) & 6.9 (+19.8)& 8.7 (+17.0)& 10.4 (+15.1)& 12.3 (+15.8)\\
4500 & 5.2 (+12.8) & 8.5 (+16.7)& 10.6 (+14.5)& 11.2 (+13.7)& 11.1 (+12.5)& 11.8 (+13.9)\\
4000 & 7.0 (+16.4)& 8.6 (+12.9)& 9.6 (+11.9)& 10.1 (+11.7)& 10.1 (+11.2)& 11.1 (+12.9)\\
3500 & 6.6 (+13.5)& 7.9 (+11.1)& 8.8 (+10.6)& 9.4 (+10.6) & 9.6 (+10.5) & 10.7 (+12.4)\\
3000 & 6.1 (+11.5)& 7.3 (+9.8)& 8.2 (+9.6)& 8.9 (+9.9)& 9.2 (+10.0)& 10.4 (+12.0)
\enddata
\tablecomments{Numbers without parentheses represent the baseline systematic uncertainty due to uncertainties on the composition of the star and conductive opacities. Numbers in parentheses indicate the additional potential relative age increase due to $^{22}$Ne shell distillation.}
\end{deluxetable*}
\vspace{-20pt}

To better understand the relative contributions of the different uncertainty sources, we performed linear regression analyses on ages versus various parameters for a 0.6$\,M_{\odot}$ white dwarf at 4000\,K. This analysis reveals that across the full range of parameters explored, (1) variations in the total oxygen content of the core lead to age differences of up to 0.8\,Gyr, (2) changes in hydrogen layer thickness result in age variations of up to 1.0\,Gyr, (3) helium layer thickness contributes up to 0.6\,Gyr in age differences, and (4) uncertainties in conductive opacities account for up to 1.3\,Gyr in age variations (see \citealt{Salaris2022} for a similar result on conductive opacities). Note that these values represent the maximum variations across the entire parameter range, effectively corresponding to more than 1$\sigma$ uncertainties. This is why simply summing these numbers (in quadrature) yields an uncertainty larger than the corresponding 1$\sigma$ uncertainty shown in Figure~\ref{fig:age_uncert}.

We also performed an additional analysis where we excluded the asteroseismological results from \cite{Giammichele2022} when defining the parameter space for the core oxygen profile. Using only the theoretical stellar evolution models to inform our parameter distributions yielded results very similar to those shown in Figure~\ref{fig:age_uncert} (see Figure~\ref{fig:age_uncert_noastero}). Our uncertainty estimates are therefore not significantly impacted by the inclusion or exclusion of asteroseismological constraints on the core composition, which is consistent with our finding that the core composition is not the dominant source of uncertainty in white dwarf cooling ages.

\subsection{Uncertainties with distillation}
We now present results that incorporate the potential impact of $^{22}$Ne shell distillation on white dwarf cooling ages. As discussed in Section \ref{sec:methods}, this process can only increase cooling timescales compared to currently available cooling tracks \citep{Renedo2010,Bedard2020,Salaris2022}. Therefore, the age uncertainty induced by shell distillation cannot be reported as a symmetric spread around a mean value, since this effect exclusively pushes cooling ages in one direction---towards older ages. Accordingly, we describe the uncertainty resulting from the possible occurrence of shell distillation in the general white dwarf population as a potential shift in cooling timescales. This approach is reflected in Table~\ref{tab:rel_age_uncert}.

To interpret Table~\ref{tab:rel_age_uncert}, it is useful to consider a few examples. For a 0.8$\,M_\odot$ white dwarf at $20{,}000\,$K, where no distillation can possibly have taken place because core crystallization has yet to begin, we report a simple systematic uncertainty of 2.6\%. However, at 6000\,K and below, distillation could have introduced an additional cooling delay. In these cases, we report two values: the first represents the baseline systematic uncertainty, while the value in parentheses indicates the potential additional delay due to distillation. The amplitude of this delay initially increases sharply with decreasing $T_\mathrm{eff}$, reflecting the fact that the $T_\mathrm{eff}$ range in which we suspect shell distillation to be taking place is quite narrow. In relative terms, this shift then slowly decreases at lower temperatures, since the cooling age continues to increase while the distillation delay remains constant. The corresponding absolute uncertainties in Gyr are provided in Table~\ref{tab:age_uncert}.

To illustrate the application of these results, consider a 0.8\,$M_\odot$ white dwarf at 5000\,K. Ignoring distillation, its cooling age probability density function would be represented by a Gaussian centered on the nominal cooling age derived from standard models, with a standard deviation of 8.7\% of that age. However, if shell distillation did take place, the age probability density function would be shifted by 15.6\% compared to the nominal age. We recommend always including at least the no-distillation uncertainty in age estimates, summing this uncertainty in quadrature with the observational uncertainty. However, for most applications, we strongly advise considering the possibility that distillation has further delayed the cooling. Our understanding of $^{22}$Ne distillation and its prevalence in the general white dwarf population is rapidly evolving. This source of uncertainty will hopefully be eliminated in the not-so-distant future. If that is the case, the no-distillation systematic uncertainties in Table~\ref{tab:rel_age_uncert} would remain relevant, while the potential distillation delays could be disregarded.

\section{Conclusions}
\label{sec:conclusion}
In this study, we have quantified the systematic uncertainties in white dwarf cooling age determinations arising from model uncertainties. Using \texttt{MESA}, we generated cooling sequences for hydrogen-atmosphere white dwarfs with carbon--oxygen cores, spanning masses from 0.5 to 1.0\,$M_{\odot}$. We systematically varied key parameters including the core composition profile, envelope masses, and conductive opacities within ranges informed by current stellar evolution models and asteroseismological studies. Additionally, we accounted for the potential impact of $^{22}$Ne shell distillation, a process that may significantly delay the cooling of most white dwarfs. Our results, summarized in Table~\ref{tab:rel_age_uncert}, provide estimates of the systematic uncertainties in cooling ages as a function of white dwarf mass and effective temperature. We strongly recommend that these uncertainties be incorporated into all future studies involving white dwarf cooling ages.

Our analysis reveals that the dominant sources of uncertainty in white dwarf cooling ages arise from the physics of the cooling process itself. Two key factors stand out: the potential occurrence of $^{22}$Ne shell distillation in the majority of white dwarfs and the treatment of conductive opacities in the regime of partial degeneracy and moderate coupling. While less prominent, uncertainties stemming from our incomplete understanding of earlier stellar evolution phases, which affect the composition profiles of white dwarfs, are still significant

It is important to note that our analysis focused solely on systematic uncertainties stemming from cooling calculations. Another crucial aspect not addressed in this study is the uncertainty in the physics of model atmospheres used to derive the effective temperature and mass of a white dwarf in the first place. These uncertainties can be large. For instance, there is a well-documented ``low-mass problem'' observed in white dwarfs cooler than 6000\,K, where current atmospheric models yield unexpectedly low masses \citep{Caron2023,O'Brien2024}. This issue affects approximately 45\% of the white dwarf population. Therefore, even when combining the systematic cooling uncertainties derived here with observational errors, we are likely still underestimating the true uncertainty in white dwarf age determinations.

\section*{acknowledgments}
We thank the anonymous reviewer for useful comments. \texttt{Claude 3.5 Sonnet} and \texttt{ChatGPT} were used to improve wording at the sentence level and assist with coding. The \texttt{MESA} computations presented here were performed on the Niagara supercomputer at the SciNet HPC Consortium. SciNet is funded by Innovation, Science and Economic Development Canada; the Digital Research Alliance of Canada; the Ontario Research Fund: Research Excellence; and the University of Toronto. The data analysis was carried out on the Astrohub virtual research environment (\url{https://astrohub.uvic.ca/}) hosted on the Digital Research Alliance of Canada Arbutus Cloud at the University of Victoria.

\vspace{5mm}
\software{\texttt{MESA}}

\bibliography{references}{}

\begin{thebibliography}{}
\expandafter\ifx\csname natexlab\endcsname\relax\def\natexlab#1{#1}\fi
\providecommand{\url}[1]{\href{#1}{#1}}
\providecommand{\dodoi}[1]{doi:~\href{http://doi.org/#1}{\nolinkurl{#1}}}
\providecommand{\doeprint}[1]{\href{http://ascl.net/#1}{\nolinkurl{http://ascl.net/#1}}}
\providecommand{\doarXiv}[1]{\href{https://arxiv.org/abs/#1}{\nolinkurl{https://arxiv.org/abs/#1}}}

\bibitem[{{Althaus} {et~al.}(2015){Althaus}, {Camisassa}, {Miller Bertolami}, {C{\'o}rsico}, \& {Garc{\'\i}a-Berro}}]{Althaus2015}
{Althaus}, L.~G., {Camisassa}, M.~E., {Miller Bertolami}, M.~M., {C{\'o}rsico}, A.~H., \& {Garc{\'\i}a-Berro}, E. 2015, \aap, 576, A9, \dodoi{10.1051/0004-6361/201424922}

\bibitem[{{Althaus} {et~al.}(2010){Althaus}, {C{\'o}rsico}, {Isern}, \& {Garc{\'\i}a-Berro}}]{Althaus2010}
{Althaus}, L.~G., {C{\'o}rsico}, A.~H., {Isern}, J., \& {Garc{\'\i}a-Berro}, E. 2010, \aapr, 18, 471, \dodoi{10.1007/s00159-010-0033-1}

\bibitem[{{Althaus} {et~al.}(2012){Althaus}, {Garc{\'\i}a-Berro}, {Isern}, {C{\'o}rsico}, \& {Miller Bertolami}}]{Althaus2012}
{Althaus}, L.~G., {Garc{\'\i}a-Berro}, E., {Isern}, J., {C{\'o}rsico}, A.~H., \& {Miller Bertolami}, M.~M. 2012, \aap, 537, A33, \dodoi{10.1051/0004-6361/201117902}

\bibitem[{{Althaus} {et~al.}(2005{\natexlab{a}}){Althaus}, {Miller Bertolami}, {C{\'o}rsico}, {Garc{\'\i}a-Berro}, \& {Gil-Pons}}]{Althaus2005b}
{Althaus}, L.~G., {Miller Bertolami}, M.~M., {C{\'o}rsico}, A.~H., {Garc{\'\i}a-Berro}, E., \& {Gil-Pons}, P. 2005{\natexlab{a}}, \aap, 440, L1, \dodoi{10.1051/0004-6361:200500159}

\bibitem[{{Althaus} {et~al.}(2005{\natexlab{b}}){Althaus}, {Serenelli}, {Panei}, {C{\'o}rsico}, {Garc{\'\i}a-Berro}, \& {Sc{\'o}ccola}}]{Althaus2005}
{Althaus}, L.~G., {Serenelli}, A.~M., {Panei}, J.~A., {et~al.} 2005{\natexlab{b}}, \aap, 435, 631, \dodoi{10.1051/0004-6361:20041965}

\bibitem[{{Barrientos} {et~al.}(2024){Barrientos}, {Kilic}, {Bergeron}, {Blouin}, {Brown}, \& {Andrews}}]{Barrientos2024}
{Barrientos}, M., {Kilic}, M., {Bergeron}, P., {et~al.} 2024, arXiv e-prints, arXiv:2407.18763, \dodoi{10.48550/arXiv.2407.18763}

\bibitem[{{Bauer}(2023)}]{Bauer2023}
{Bauer}, E.~B. 2023, \apj, 950, 115, \dodoi{10.3847/1538-4357/acd057}

\bibitem[{{B{\'e}dard}(2024)}]{Bedard2024}
{B{\'e}dard}, A. 2024, \apss, 369, 43, \dodoi{10.1007/s10509-024-04307-5}

\bibitem[{{B{\'e}dard} {et~al.}(2020){B{\'e}dard}, {Bergeron}, {Brassard}, \& {Fontaine}}]{Bedard2020}
{B{\'e}dard}, A., {Bergeron}, P., {Brassard}, P., \& {Fontaine}, G. 2020, \apj, 901, 93, \dodoi{10.3847/1538-4357/abafbe}

\bibitem[{{B{\'e}dard} {et~al.}(2024){B{\'e}dard}, {Blouin}, \& {Cheng}}]{Bedard2024N}
{B{\'e}dard}, A., {Blouin}, S., \& {Cheng}, S. 2024, \nat, 627, 286, \dodoi{10.1038/s41586-024-07102-y}

\bibitem[{{Blouin} {et~al.}(2021){Blouin}, {Daligault}, \& {Saumon}}]{Blouin2021L}
{Blouin}, S., {Daligault}, J., \& {Saumon}, D. 2021, \apjl, 911, L5, \dodoi{10.3847/2041-8213/abf14b}

\bibitem[{{Blouin} {et~al.}(2020{\natexlab{a}}){Blouin}, {Daligault}, {Saumon}, {B{\'e}dard}, \& {Brassard}}]{Blouin2020b}
{Blouin}, S., {Daligault}, J., {Saumon}, D., {B{\'e}dard}, A., \& {Brassard}, P. 2020{\natexlab{a}}, \aap, 640, L11, \dodoi{10.1051/0004-6361/202038879}

\bibitem[{{Blouin} {et~al.}(2024){Blouin}, {Herwig}, {Mao}, {Denissenkov}, \& {Woodward}}]{Blouin2024}
{Blouin}, S., {Herwig}, F., {Mao}, H., {Denissenkov}, P., \& {Woodward}, P.~R. 2024, \mnras, 527, 4847, \dodoi{10.1093/mnras/stad3518}

\bibitem[{{Blouin} {et~al.}(2020{\natexlab{b}}){Blouin}, {Shaffer}, {Saumon}, \& {Starrett}}]{Blouin2020}
{Blouin}, S., {Shaffer}, N.~R., {Saumon}, D., \& {Starrett}, C.~E. 2020{\natexlab{b}}, \apj, 899, 46, \dodoi{10.3847/1538-4357/ab9e75}

\bibitem[{{Blouin} \& {Xu}(2022)}]{Blouin2022}
{Blouin}, S., \& {Xu}, S. 2022, \mnras, 510, 1059, \dodoi{10.1093/mnras/stab3446}

\bibitem[{{Camisassa} {et~al.}(2019){Camisassa}, {Althaus}, {C{\'o}rsico}, {De Ger{\'o}nimo}, {Miller Bertolami}, {Novarino}, {Rohrmann}, {Wachlin}, \& {Garc{\'\i}a-Berro}}]{Camisassa2019}
{Camisassa}, M.~E., {Althaus}, L.~G., {C{\'o}rsico}, A.~H., {et~al.} 2019, \aap, 625, A87, \dodoi{10.1051/0004-6361/201833822}

\bibitem[{{Caron} {et~al.}(2023){Caron}, {Bergeron}, {Blouin}, \& {Leggett}}]{Caron2023}
{Caron}, A., {Bergeron}, P., {Blouin}, S., \& {Leggett}, S.~K. 2023, \mnras, 519, 4529, \dodoi{10.1093/mnras/stac3733}

\bibitem[{{Cassisi} {et~al.}(2007){Cassisi}, {Potekhin}, {Pietrinferni}, {Catelan}, \& {Salaris}}]{Cassisi2007}
{Cassisi}, S., {Potekhin}, A.~Y., {Pietrinferni}, A., {Catelan}, M., \& {Salaris}, M. 2007, \apj, 661, 1094, \dodoi{10.1086/516819}

\bibitem[{{Cassisi} {et~al.}(2021){Cassisi}, {Potekhin}, {Salaris}, \& {Pietrinferni}}]{Cassisi2021}
{Cassisi}, S., {Potekhin}, A.~Y., {Salaris}, M., \& {Pietrinferni}, A. 2021, \aap, 654, A149, \dodoi{10.1051/0004-6361/202141425}

\bibitem[{{Castanheira} \& {Kepler}(2009)}]{Castanheira2009}
{Castanheira}, B.~G., \& {Kepler}, S.~O. 2009, \mnras, 396, 1709, \dodoi{10.1111/j.1365-2966.2009.14855.x}

\bibitem[{{Cheng} {et~al.}(2019){Cheng}, {Cummings}, \& {M{\'e}nard}}]{Cheng2019}
{Cheng}, S., {Cummings}, J.~D., \& {M{\'e}nard}, B. 2019, \apj, 886, 100, \dodoi{10.3847/1538-4357/ab4989}

\bibitem[{{Chidester} \& {Timmes}(2022)}]{Chidester2022}
{Chidester}, M., \& {Timmes}, F. 2022, in American Astronomical Society Meeting Abstracts, Vol. 240, American Astronomical Society Meeting \#240, 415.05

\bibitem[{{Cimatti} \& {Moresco}(2023)}]{Cimatti2023}
{Cimatti}, A., \& {Moresco}, M. 2023, \apj, 953, 149, \dodoi{10.3847/1538-4357/ace439}

\bibitem[{{Constantino} {et~al.}(2015){Constantino}, {Campbell}, {Christensen-Dalsgaard}, {Lattanzio}, \& {Stello}}]{Constantino2015}
{Constantino}, T., {Campbell}, S.~W., {Christensen-Dalsgaard}, J., {Lattanzio}, J.~C., \& {Stello}, D. 2015, \mnras, 452, 123, \dodoi{10.1093/mnras/stv1264}

\bibitem[{{Cukanovaite} {et~al.}(2023){Cukanovaite}, {Tremblay}, {Toonen}, {Temmink}, {Manser}, {O'Brien}, \& {McCleery}}]{Cukanovaite2023}
{Cukanovaite}, E., {Tremblay}, P.~E., {Toonen}, S., {et~al.} 2023, \mnras, 522, 1643, \dodoi{10.1093/mnras/stad1020}

\bibitem[{{Cummings} {et~al.}(2018){Cummings}, {Kalirai}, {Tremblay}, {Ramirez-Ruiz}, \& {Choi}}]{Cummings2018}
{Cummings}, J.~D., {Kalirai}, J.~S., {Tremblay}, P.~E., {Ramirez-Ruiz}, E., \& {Choi}, J. 2018, \apj, 866, 21, \dodoi{10.3847/1538-4357/aadfd6}

\bibitem[{{Cunningham} {et~al.}(2024){Cunningham}, {Tremblay}, \& {W. O'Brien}}]{Cunningham2024}
{Cunningham}, T., {Tremblay}, P.-E., \& {W. O'Brien}, M. 2024, \mnras, 527, 3602, \dodoi{10.1093/mnras/stad3275}

\bibitem[{{De Ger{\'o}nimo} {et~al.}(2017){De Ger{\'o}nimo}, {Althaus}, {C{\'o}rsico}, {Romero}, \& {Kepler}}]{DeGeronimo2017}
{De Ger{\'o}nimo}, F.~C., {Althaus}, L.~G., {C{\'o}rsico}, A.~H., {Romero}, A.~D., \& {Kepler}, S.~O. 2017, \aap, 599, A21, \dodoi{10.1051/0004-6361/201629806}

\bibitem[{{Elms} {et~al.}(2022){Elms}, {Tremblay}, {G{\"a}nsicke}, {Koester}, {Hollands}, {Gentile Fusillo}, {Cunningham}, \& {Apps}}]{Elms2022}
{Elms}, A.~K., {Tremblay}, P.-E., {G{\"a}nsicke}, B.~T., {et~al.} 2022, \mnras, 517, 4557, \dodoi{10.1093/mnras/stac2908}

\bibitem[{{Fantin} {et~al.}(2019){Fantin}, {C{\^o}t{\'e}}, {McConnachie}, {Bergeron}, {Cuillandre}, {Gwyn}, {Ibata}, {Thomas}, {Carlberg}, {Fabbro}, {Haywood}, {Lan{\c{c}}on}, {Lewis}, {Malhan}, {Martin}, {Navarro}, {Scott}, \& {Starkenburg}}]{Fantin2019}
{Fantin}, N.~J., {C{\^o}t{\'e}}, P., {McConnachie}, A.~W., {et~al.} 2019, \apj, 887, 148, \dodoi{10.3847/1538-4357/ab5521}

\bibitem[{{Fontaine} {et~al.}(2001){Fontaine}, {Brassard}, \& {Bergeron}}]{Fontaine2001}
{Fontaine}, G., {Brassard}, P., \& {Bergeron}, P. 2001, \pasp, 113, 409, \dodoi{10.1086/319535}

\bibitem[{{Gaia Collaboration} {et~al.}(2018){Gaia Collaboration}, {Babusiaux}, {van Leeuwen}, {Barstow}, {Jordi}, {Vallenari}, {Bossini}, {Bressan}, {Cantat-Gaudin}, {van Leeuwen}, \& et~al.}]{GaiaCollaboration2018}
{Gaia Collaboration}, {Babusiaux}, C., {van Leeuwen}, F., {et~al.} 2018, \aap, 616, A10, \dodoi{10.1051/0004-6361/201832843}

\bibitem[{{Garc{\'\i}a-Berro} {et~al.}(2010){Garc{\'\i}a-Berro}, {Torres}, {Althaus}, {Renedo}, {Lor{\'e}n-Aguilar}, {C{\'o}rsico}, {Rohrmann}, {Salaris}, \& {Isern}}]{Garcia-Berro2010}
{Garc{\'\i}a-Berro}, E., {Torres}, S., {Althaus}, L.~G., {et~al.} 2010, \nat, 465, 194, \dodoi{10.1038/nature09045}

\bibitem[{{Giammichele} {et~al.}(2022){Giammichele}, {Charpinet}, \& {Brassard}}]{Giammichele2022}
{Giammichele}, N., {Charpinet}, S., \& {Brassard}, P. 2022, Frontiers in Astronomy and Space Sciences, 9, 879045, \dodoi{10.3389/fspas.2022.879045}

\bibitem[{{Giammichele} {et~al.}(2017){Giammichele}, {Charpinet}, {Fontaine}, \& {Brassard}}]{Giammichele2017}
{Giammichele}, N., {Charpinet}, S., {Fontaine}, G., \& {Brassard}, P. 2017, \apj, 834, 136, \dodoi{10.3847/1538-4357/834/2/136}

\bibitem[{{Herwig}(2000)}]{Herwig2000}
{Herwig}, F. 2000, \aap, 360, 952, \dodoi{10.48550/arXiv.astro-ph/0007139}

\bibitem[{{Hollands} {et~al.}(2018){Hollands}, {G{\"a}nsicke}, \& {Koester}}]{Hollands2018}
{Hollands}, M.~A., {G{\"a}nsicke}, B.~T., \& {Koester}, D. 2018, \mnras, 477, 93, \dodoi{10.1093/mnras/sty592}

\bibitem[{{Hollands} {et~al.}(2024){Hollands}, {Littlefair}, \& {Parsons}}]{Hollands2024}
{Hollands}, M.~A., {Littlefair}, S.~P., \& {Parsons}, S.~G. 2024, \mnras, 527, 9061, \dodoi{10.1093/mnras/stad3729}

\bibitem[{{Isern}(2019)}]{Isern2019}
{Isern}, J. 2019, \apjl, 878, L11, \dodoi{10.3847/2041-8213/ab238e}

\bibitem[{{Isern} {et~al.}(1991){Isern}, {Hernanz}, {Mochkovitch}, \& {Garcia-Berro}}]{Isern1991}
{Isern}, J., {Hernanz}, M., {Mochkovitch}, R., \& {Garcia-Berro}, E. 1991, \aap, 241, L29

\bibitem[{{Kaiser} {et~al.}(2021){Kaiser}, {Clemens}, {Blouin}, {Dufour}, {Hegedus}, {Reding}, \& {B{\'e}dard}}]{Kaiser2021}
{Kaiser}, B.~C., {Clemens}, J.~C., {Blouin}, S., {et~al.} 2021, Science, 371, 168, \dodoi{10.1126/science.abd1714}

\bibitem[{Kalirai(2012)}]{Kalirai2012}
Kalirai, J.~S. 2012, Nature, 486, 90, \dodoi{10.1038/nature11062}

\bibitem[{{Kilic} {et~al.}(2024){Kilic}, {Bergeron}, {Blouin}, {Moss}, {Brown}, {Bedard}, {Jewett}, \& {Agueros}}]{Kilic2024}
{Kilic}, M., {Bergeron}, P., {Blouin}, S., {et~al.} 2024, arXiv e-prints, arXiv:2412.04611, \dodoi{10.48550/arXiv.2412.04611}

\bibitem[{{Kilic} {et~al.}(2017){Kilic}, {Munn}, {Harris}, {von Hippel}, {Liebert}, {Williams}, {Jeffery}, \& {DeGennaro}}]{Kilic2017}
{Kilic}, M., {Munn}, J.~A., {Harris}, H.~C., {et~al.} 2017, \apj, 837, 162, \dodoi{10.3847/1538-4357/aa62a5}

\bibitem[{{Kiman} {et~al.}(2022){Kiman}, {Xu}, {Faherty}, {Gagn{\'e}}, {Angus}, {Brandt}, {Casewell}, \& {Cruz}}]{Kiman2022}
{Kiman}, R., {Xu}, S., {Faherty}, J.~K., {et~al.} 2022, \aj, 164, 62, \dodoi{10.3847/1538-3881/ac7788}

\bibitem[{{Mestel}(1952)}]{Mestel1952}
{Mestel}, L. 1952, \mnras, 112, 583, \dodoi{10.1093/mnras/112.6.583}

\bibitem[{{Moss} {et~al.}(2022){Moss}, {von Hippel}, {Robinson}, {El-Badry}, {Stenning}, {van Dyk}, {Fouesneau}, {Bailer-Jones}, {Jeffery}, {Sargent}, {Kloc}, \& {Moticska}}]{Moss2022}
{Moss}, A., {von Hippel}, T., {Robinson}, E., {et~al.} 2022, \apj, 929, 26, \dodoi{10.3847/1538-4357/ac5ac0}

\bibitem[{{O'Brien} {et~al.}(2024){O'Brien}, {Tremblay}, {Klein}, {Koester}, {Melis}, {B{\'e}dard}, {Cukanovaite}, {Cunningham}, {Doyle}, {G{\"a}nsicke}, {Gentile Fusillo}, {Hollands}, {McCleery}, {Pelisoli}, {Toonen}, {Weinberger}, \& {Zuckerman}}]{O'Brien2024}
{O'Brien}, M.~W., {Tremblay}, P.~E., {Klein}, B.~L., {et~al.} 2024, \mnras, 527, 8687, \dodoi{10.1093/mnras/stad3773}

\bibitem[{{Paxton} {et~al.}(2011){Paxton}, {Bildsten}, {Dotter}, {Herwig}, {Lesaffre}, \& {Timmes}}]{MESA1}
{Paxton}, B., {Bildsten}, L., {Dotter}, A., {et~al.} 2011, \apjs, 192, 3, \dodoi{10.1088/0067-0049/192/1/3}

\bibitem[{{Paxton} {et~al.}(2013){Paxton}, {Cantiello}, {Arras}, {Bildsten}, {Brown}, {Dotter}, {Mankovich}, {Montgomery}, {Stello}, {Timmes}, \& {Townsend}}]{MESA2}
{Paxton}, B., {Cantiello}, M., {Arras}, P., {et~al.} 2013, \apjs, 208, 4, \dodoi{10.1088/0067-0049/208/1/4}

\bibitem[{{Paxton} {et~al.}(2015){Paxton}, {Marchant}, {Schwab}, {Bauer}, {Bildsten}, {Cantiello}, {Dessart}, {Farmer}, {Hu}, {Langer}, {Townsend}, {Townsley}, \& {Timmes}}]{MESA3}
{Paxton}, B., {Marchant}, P., {Schwab}, J., {et~al.} 2015, \apjs, 220, 15, \dodoi{10.1088/0067-0049/220/1/15}

\bibitem[{{Paxton} {et~al.}(2018){Paxton}, {Schwab}, {Bauer}, {Bildsten}, {Blinnikov}, {Duffell}, {Farmer}, {Goldberg}, {Marchant}, {Sorokina}, {Thoul}, {Townsend}, \& {Timmes}}]{MESA4}
{Paxton}, B., {Schwab}, J., {Bauer}, E.~B., {et~al.} 2018, \apjs, 234, 34, \dodoi{10.3847/1538-4365/aaa5a8}

\bibitem[{{Paxton} {et~al.}(2019){Paxton}, {Smolec}, {Schwab}, {Gautschy}, {Bildsten}, {Cantiello}, {Dotter}, {Farmer}, {Goldberg}, {Jermyn}, {Kanbur}, {Marchant}, {Thoul}, {Townsend}, {Wolf}, {Zhang}, \& {Timmes}}]{MESA5}
{Paxton}, B., {Smolec}, R., {Schwab}, J., {et~al.} 2019, \apjs, 243, 10, \dodoi{10.3847/1538-4365/ab2241}

\bibitem[{{Pepper} {et~al.}(2022){Pepper}, {Istrate}, {Romero}, \& {Kepler}}]{Pepper2022}
{Pepper}, B.~T., {Istrate}, A.~G., {Romero}, A.~D., \& {Kepler}, S.~O. 2022, \mnras, 513, 1499, \dodoi{10.1093/mnras/stac1016}

\bibitem[{{Renedo} {et~al.}(2010){Renedo}, {Althaus}, {Miller Bertolami}, {Romero}, {C{\'o}rsico}, {Rohrmann}, \& {Garc{\'\i}a-Berro}}]{Renedo2010}
{Renedo}, I., {Althaus}, L.~G., {Miller Bertolami}, M.~M., {et~al.} 2010, \apj, 717, 183, \dodoi{10.1088/0004-637X/717/1/183}

\bibitem[{{Rohrmann} {et~al.}(2011){Rohrmann}, {Althaus}, \& {Kepler}}]{Rohrmann2011}
{Rohrmann}, R.~D., {Althaus}, L.~G., \& {Kepler}, S.~O. 2011, \mnras, 411, 781, \dodoi{10.1111/j.1365-2966.2010.17716.x}

\bibitem[{{Rolland} {et~al.}(2018){Rolland}, {Bergeron}, \& {Fontaine}}]{Rolland2018}
{Rolland}, B., {Bergeron}, P., \& {Fontaine}, G. 2018, \apj, 857, 56, \dodoi{10.3847/1538-4357/aab713}

\bibitem[{{Romero} {et~al.}(2013){Romero}, {Kepler}, {C{\'o}rsico}, {Althaus}, \& {Fraga}}]{Romero2013}
{Romero}, A.~D., {Kepler}, S.~O., {C{\'o}rsico}, A.~H., {Althaus}, L.~G., \& {Fraga}, L. 2013, \apj, 779, 58, \dodoi{10.1088/0004-637X/779/1/58}

\bibitem[{{Salaris} {et~al.}(2024){Salaris}, {Blouin}, {Cassisi}, \& {Bedin}}]{Salaris2024}
{Salaris}, M., {Blouin}, S., {Cassisi}, S., \& {Bedin}, L.~R. 2024, arXiv e-prints, arXiv:2403.02790, \dodoi{10.48550/arXiv.2403.02790}

\bibitem[{{Salaris} \& {Cassisi}(2017)}]{Salaris2017}
{Salaris}, M., \& {Cassisi}, S. 2017, Royal Society Open Science, 4, 170192, \dodoi{10.1098/rsos.170192}

\bibitem[{{Salaris} {et~al.}(2022){Salaris}, {Cassisi}, {Pietrinferni}, \& {Hidalgo}}]{Salaris2022}
{Salaris}, M., {Cassisi}, S., {Pietrinferni}, A., \& {Hidalgo}, S. 2022, \mnras, 509, 5197, \dodoi{10.1093/mnras/stab3359}

\bibitem[{{Salaris} {et~al.}(1997){Salaris}, {Dom{\'\i}nguez}, {Garc{\'\i}a-Berro}, {Hernanz}, {Isern}, \& {Mochkovitch}}]{Salaris1997}
{Salaris}, M., {Dom{\'\i}nguez}, I., {Garc{\'\i}a-Berro}, E., {et~al.} 1997, \apj, 486, 413, \dodoi{10.1086/304483}

\bibitem[{{Saumon} {et~al.}(2022){Saumon}, {Blouin}, \& {Tremblay}}]{Saumon2022}
{Saumon}, D., {Blouin}, S., \& {Tremblay}, P.-E. 2022, \physrep, 988, 1, \dodoi{10.1016/j.physrep.2022.09.001}

\bibitem[{{Segretain}(1996)}]{Segretain1996}
{Segretain}, L. 1996, \aap, 310, 485, \dodoi{10.48550/arXiv.astro-ph/9510118}

\bibitem[{{Shen} {et~al.}(2023){Shen}, {Blouin}, \& {Breivik}}]{Shen2023}
{Shen}, K.~J., {Blouin}, S., \& {Breivik}, K. 2023, \apjl, 955, L33, \dodoi{10.3847/2041-8213/acf57b}

\bibitem[{{Straniero} {et~al.}(2003){Straniero}, {Dom{\'\i}nguez}, {Imbriani}, \& {Piersanti}}]{Straniero2003}
{Straniero}, O., {Dom{\'\i}nguez}, I., {Imbriani}, G., \& {Piersanti}, L. 2003, \apj, 583, 878, \dodoi{10.1086/345427}

\bibitem[{{Tremblay} {et~al.}(2019){Tremblay}, {Fontaine}, {Gentile Fusillo}, {Dunlap}, {G{\"a}nsicke}, {Hollands}, {Hermes}, {Marsh}, {Cukanovaite}, \& {Cunningham}}]{Tremblay2019}
{Tremblay}, P.-E., {Fontaine}, G., {Gentile Fusillo}, N.~P., {et~al.} 2019, \nat, 565, 202, \dodoi{10.1038/s41586-018-0791-x}

\bibitem[{{Venner} {et~al.}(2023){Venner}, {Blouin}, {B{\'e}dard}, \& {Vanderburg}}]{Venner2023}
{Venner}, A., {Blouin}, S., {B{\'e}dard}, A., \& {Vanderburg}, A. 2023, \mnras, 523, 4624, \dodoi{10.1093/mnras/stad1719}

\bibitem[{{Weiss} \& {Ferguson}(2009)}]{Weiss2009}
{Weiss}, A., \& {Ferguson}, J.~W. 2009, \aap, 508, 1343, \dodoi{10.1051/0004-6361/200912043}

\bibitem[{{Winget} {et~al.}(1987){Winget}, {Hansen}, {Liebert}, {van Horn}, {Fontaine}, {Nather}, {Kepler}, \& {Lamb}}]{Winget1987}
{Winget}, D.~E., {Hansen}, C.~J., {Liebert}, J., {et~al.} 1987, \apjl, 315, L77, \dodoi{10.1086/184864}

\end{thebibliography}
\bibliographystyle{aasjournal}

\clearpage

\appendix
\FloatBarrier
\renewcommand{\thefigure}{A\arabic{figure}}
\renewcommand{\thetable}{A\arabic{table}}
\setcounter{figure}{0}
\setcounter{table}{0}

\begin{figure*}
\centering
\includegraphics[width=0.49\textwidth]{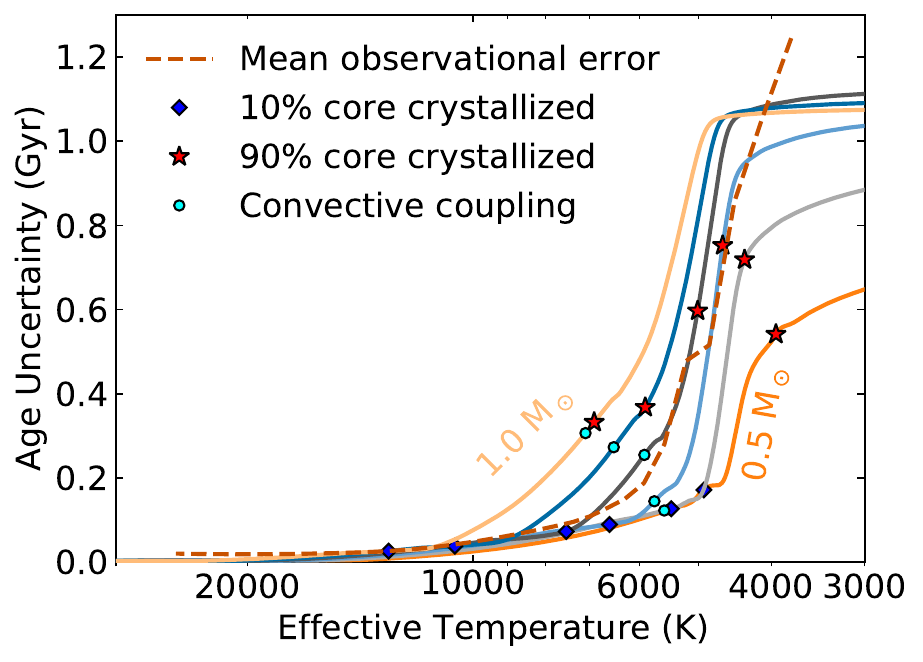}
\includegraphics[width=0.49\textwidth]{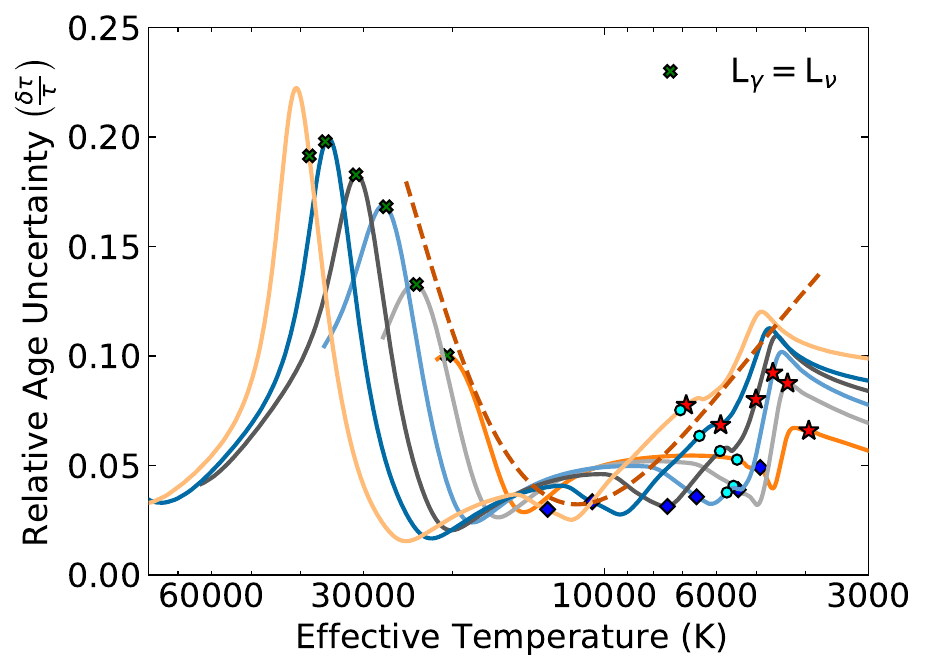}
\caption{Same as Figure~\ref{fig:age_uncert}, but using only theoretical stellar evolution models to define the parameter space for the core oxygen profile.}
\label{fig:age_uncert_noastero}
\end{figure*}

\begin{deluxetable*}{ccccccc}
\tablecaption{Systematic cooling age uncertainty in Gyr as a function of mass and effective temperature for hydrogen-atmosphere white dwarfs with carbon--oxygen cores
\label{tab:age_uncert}}
\tablewidth{0pt}
\tablehead{
\colhead{$T_\mathrm{eff}$ (K)} & 
\colhead{0.5\,$M_\odot$} & 
\colhead{0.6\,$M_\odot$} & 
\colhead{0.7\,$M_\odot$} & 
\colhead{0.8\,$M_\odot$} & 
\colhead{0.9\,$M_\odot$} & 
\colhead{1.0\,$M_\odot$}
}
\startdata
20000 & 0.00  & 0.00 & 0.00 & 0.00 & 0.00 & 0.01 \\
17500 & 0.00  & 0.00 & 0.01 & 0.01 & 0.01 & 0.01 \\
15000 & 0.00  & 0.01 & 0.01 & 0.01 & 0.02 & 0.02 \\
12500 & 0.01  & 0.02 & 0.02 & 0.03 & 0.03 & 0.03 \\
10000 & 0.03  & 0.03 & 0.04 & 0.05 & 0.05 & 0.10 (+0.72) \\
9000 & 0.03  & 0.04 & 0.05 & 0.06 & 0.06 (+0.18)& 0.16 (+1.26) \\
8000 & 0.05  & 0.06 & 0.07 & 0.07 & 0.14 (+0.87)& 0.24 (+1.26) \\
7000 & 0.07  & 0.08 & 0.10 & 0.12 & 0.25 (+1.24)& 0.35 (+1.27)\\
6000 & 0.10  & 0.12 & 0.13 & 0.29 (+0.88)& 0.39 (+1.24)& 0.51 (+1.27)\\
5500 & 0.12  & 0.14 & 0.22 (+0.47)& 0.38 (+1.23)& 0.52 (+1.23)& 0.70 (+1.28)\\
5000 & 0.18  & 0.19 (+0.44) & 0.43 (+1.25)& 0.65 (+1.25)& 0.84 (+1.22)& 1.00 (+1.28)\\
4500 & 0.29 (+0.72) & 0.63 (+1.24)& 0.92 (+1.26)& 1.07 (+1.30)& 1.08 (+1.22)& 1.09 (+1.28)\\
4000 & 0.55 (+1.29)& 0.82 (+1.23)& 1.02 (+1.26)& 1.12 (+1.29)& 1.10 (+1.22)& 1.10 (+1.28)\\
3500 & 0.63 (+1.29)& 0.88 (+1.23)& 1.06 (+1.27)& 1.15 (+1.29) & 1.11 (+1.21) & 1.11 (+1.28)\\
3000 & 0.69 (+1.29)& 0.92 (+1.24)& 1.08 (+1.27)& 1.16 (+1.29)& 1.12 (+1.21)& 1.11 (+1.28)
\enddata
\tablecomments{Numbers without parentheses represent the baseline systematic uncertainty due to uncertainties on the composition of the star and conductive opacities. Numbers in parentheses indicate the additional potential age increase due to $^{22}$Ne shell distillation.}
\end{deluxetable*}

\end{document}